\documentclass[a4paper, amsfonts, amssymb, amsmath, reprint, showkeys, nofootinbib, twoside]{revtex4-1}
\usepackage[english]{babel}
\usepackage[utf8]{inputenc}
\usepackage[colorinlistoftodos, color=green!40, prependcaption]{todonotes}
\usepackage{amsthm}
\usepackage{mathtools}
\usepackage{physics}
\usepackage{xcolor}
\usepackage{graphicx}
\usepackage[left=23mm,right=13mm,top=35mm,columnsep=15pt]{geometry} 
\usepackage{adjustbox}
\usepackage{placeins}
\usepackage[T1]{fontenc}
\usepackage{lipsum}
\usepackage{csquotes}
\usepackage[pdftex, pdftitle={Article}, pdfauthor={Author}]{hyperref} 
\begin{document}
\title{Orthogonal Spatial Coding with Stimulated Parametric Down-Conversion}

\author{Yang Xu}
    \email[Correspondence email address: ]{yxu100@ur.rochester.edu}
    \affiliation{Department of Physics and Astronomy, University of Rochester, Rochester, New York, 14627}
\author{Sirui Tang}
    \affiliation{The Institute of Optics, University of Rochester, Rochester, New York, 14627}
\author{A. Nicholas Black}
    \affiliation{Department of Physics and Astronomy, University of Rochester, NY}
\author{Robert W. Boyd}
    \affiliation{Department of Physics and Astronomy, University of Rochester, Rochester, New York, 14627}
    \affiliation{The Institute of Optics, University of Rochester, Rochester, New York, 14627}
    \affiliation{Department of Physics, University of Ottawa, Ottawa, Ontario K1N 6N5, Canada}
    
\date{\today} 

\begin{abstract}
Orthogonal optical coding is widely used in classical multiuser communication networks. Using the phase conjugation property of stimulated parametric down-conversion, we extend the current orthogonal optical coding scheme to the spatial domain to encode and decode image information. In this process, the idler beam inherits the complex conjugate of the field information encoded in the seed beam. An encoding phase mask introduced to the input seed beam blurs the image transferred to the idler. The original image is restored by passing the coded transferred image through a corrective phase mask placed in the momentum space of the idler beam. We expect that this scheme can also inspire new techniques in aberration cancellation and frequency conversion imaging. 
\end{abstract}


\maketitle
Wavefront aberrations and group velocity dispersions (GVD) negatively degrade the image quality in both classical and quantum imaging schemes. In particular, quantum imaging schemes\cite{gi,qiup} are highly vulnerable to GVD and aberrations because the correlated photon pairs produced in the spontaneous parametric down-conversion (SPDC) become disentangled even under very small perturbations. Recently, multiple schemes have been demonstrated to correct the GVD \cite{PhysRevA.45.3126} and wavefront aberrations \cite{PhysRevLett.87.123602} in quantum imaging experiments. A local cancellation of GVD based on indistinguishability \cite{PhysRevLett.68.2421, PhysRevA.47.3291} and a nonlocal cancellation based on the frequency correlations \cite{PhysRevLett.120.053601, Baek:09, PhysRevA.88.020103} have been proposed and demonstrated. On the other hand, aberrations caused by the phase shifts of the light's transverse momentum can be alleviated in a conceptually similar way because aberration is the spatial counterpart to dispersion due to space-time duality \cite{301659}. Both local aberration cancellations \cite{PhysRevLett.101.233603, Filpi:15} and nonlocal cancellations\cite{aber} which use the position-momentum entangled photon pairs produced in a spontaneous parametric down-conversion have also been performed experimentally.  

The proven effectiveness of aberration and dispersion cancellation has prompted studies on the use of phase distortion as means to encrypt information in either the temporal or spatial domain. The orthogonal spectral coding\cite{PhysRevLett.112.133602} was first realized with the time correlation of entangled photons. A more recent work \cite{johnson2023hiding} demonstrated the possibility to fully recover an image from background noise using the temporal correlation of photon pairs produced in SPDC. The space-time duality suggests that a coding protocol of image transfer in the spatial domain should be possible in principle, but the feasibility to  implement the orthogonal optical coding with spatial correlations of entangled photons is yet to be explored.

In this letter, we demonstrate an orthogonal optical coding scheme using the spatial correlation of the seed-idler pair in a stimulated parametric down-conversion. Recent theoretical work \cite{set} has demonstrated that the stimulated parametric down-conversion, also known as the difference frequency generation (DFG), provides a more efficient method to extract the same information of the photon pair produced by its the quantum counterparts -- SPDC \cite{set_app}. This close connection between DFG and SPDC has initiated a number of new methods that are more efficient to characterize and harness the spatial correlation of photon pairs \cite{speckle, triangle}. Taking advantage of the high photon flux in the seeded process, we show that a practical image coding scheme using the spatial correlation of photon pairs can be realized through DFG. In a DFG process, the seed beam $\mathbf{E}_s$ and the pump beam $\mathbf{E}_p$ are mixed in a nonlinear crystal and generate a third beam $\mathbf{E}_i$, which we call the idler beam. Each field has a transverse spatial profile $\mathcal{E}_j(\mathbf{r})$ $(j = p, i, s)$:

\begin{align}
    \mathbf{E}_j = \mathcal{E}_j(\mathbf{r})e^{ik_jz}\mathbf{e}_j
\end{align}
where $\mathbf{e}_j (j = p, i, s)$ is the unit polarization vector. 
The process satisfies the energy conservation $\omega_p - \omega_s = \omega_i$ and the momentum conservation $\mathbf{k}_p-\mathbf{k}_s = \mathbf{k}_i$. Assuming that the pump is undepleted by the nonlinear interaction and the process is perfectly phase-matched, the coupled-amplitude equations describing the DFG are given by

\begin{align}\label{eq:idler}
    \frac{d\mathcal{E}_i}{dz} &= \frac{2id_{eff}\omega_i^2}{k_ic^2}\mathcal{E}_p\mathcal{E}^*_s
\end{align}

\begin{align}\label{eq:seed}
        \frac{d\mathcal{E}_s}{dz} &= \frac{2id_{eff}\omega_s^2}{k_sc^2}\mathcal{E}_p\mathcal{E}^*_i
\end{align}

where $d_{eff} = \frac{1}{2}\chi^{(2)}$ describes the second-order nonlinearity of the crystal.
In the low-gain regime where the thin-crystal limit holds, the spatial profile of the down-converted idler beam obtained from Equation (\ref{eq:idler}) and Equation (\ref{eq:seed}) is proportional to the product of the pump beam spatial structure and the conjugate of the seed beam spatial profile\cite{img_xfer}. 

\begin{align}\label{eq:xfer}
    \mathcal{E}_i(\mathbf{r}) \propto \mathcal{E}_p(\mathbf{r})\mathcal{E}_s^*(\mathbf{r})
\end{align}

The code used to encrypt the image is described by a momentum-dependent phase factor $e^{i\phi_s(\mathbf{k}_s)}$ acting on the angular spectrum of the seed beam that carries the image:
 
 \begin{align}
     \widetilde{\mathcal{E}}_s(\mathbf{k}_s) = \widetilde{\mathcal{E}}_{s,\text{message}}(\mathbf{k}_s) e^{i\phi_s(\mathbf{k}_s)}
 \end{align}
 
 where $ \widetilde{\mathcal{E}}_{s,\text{message}}(\mathbf{k}_s)$ is the angular spectrum of the image message encoded in the seed beam. 
 
 Using the property given by Equation \ref{eq:xfer}, we can express the angular spectrum of the down-converted idler beam 
 \begin{align}
      \widetilde{\mathcal{E}}_i(\mathbf{k}_i) \propto \widetilde{\mathcal{E}}_p(\mathbf{k}_p) \ast \widetilde{\mathcal{E}}_{s,\text{message}}^*(-\mathbf{k}_s)e^{-i\phi_s(-\mathbf{k}_s)}
 \end{align}
 where $\widetilde{\mathcal{E}}_p(\mathbf{k}_p) $ is the angular spectrum of the pump wave and $*$ is the convolution operator. Thus, under the plane-wave-pump approximation, we are able to completely filter the effect of the encoding aberrations by placing a momentum-dependent phase mask $\phi_i(\mathbf{k}_i) = \phi_s(-\mathbf{k}_s)$ at the Fourier plane of the idler arm. It must be noticed that if the pump is not a plane wave, the momenta of the seed and idler photons are not perfectly anti-correlated. This can make the complete decryption impossible to realize. 

The fundamental block to realized code division multiple access (CDMA) is a set of orthogonal codes. The so-called "Hadamard codes" are often a practical option. A given set of such codes consists of $N$ length-$N$ sequences of $1$ and $-1$, or equivalently, phase $0$ or $\pi$. This set of codes possesses the orthogonality property: any two different sequences are orthogonal. If we express any two arbitrary code sequences $m$ and $n$ in terms of vectors $\mathbf{v}_m$ and $\mathbf{v}_n$ respectively, we have 

\begin{align}
    \mathbf{v}_m \cdot \mathbf{v}_n = N\delta_{mn}
\end{align}

where $\delta_{mn}$ is the Kronecker delta, equal to $1$ if $m=n$ and $0$ otherwise.

The conventional physical realization of this protocol is the spectral coding of coherent optical pulses \cite{zheng2000spectral, zheng2001low, PhysRevLett.112.133602} where the sender encodes the message with a randomly generated code to one half of the spectrum, and only the receiver who applies the correct code to the opposite half can retrieve the image from the noise. We can extend this orthogonal spectral coding method to the spatial domain where we can apply a spatial code matrix to the angular spectrum of the image we want to send. That is, we introduce artificial aberrations by passing the Fourier transform of the image through an encoding phase mask. The phase mask takes the form of a $N \times N$ matrix whose entry is randomly chosen from two values, $0$ and $\pi$. Only the receiver who applies the same phase mask can recover the image. Otherwise, the receiver only measures a randomly distorted image. 

\begin{figure}[hbt!]
  \centering
  \includegraphics[width=0.5\textwidth]{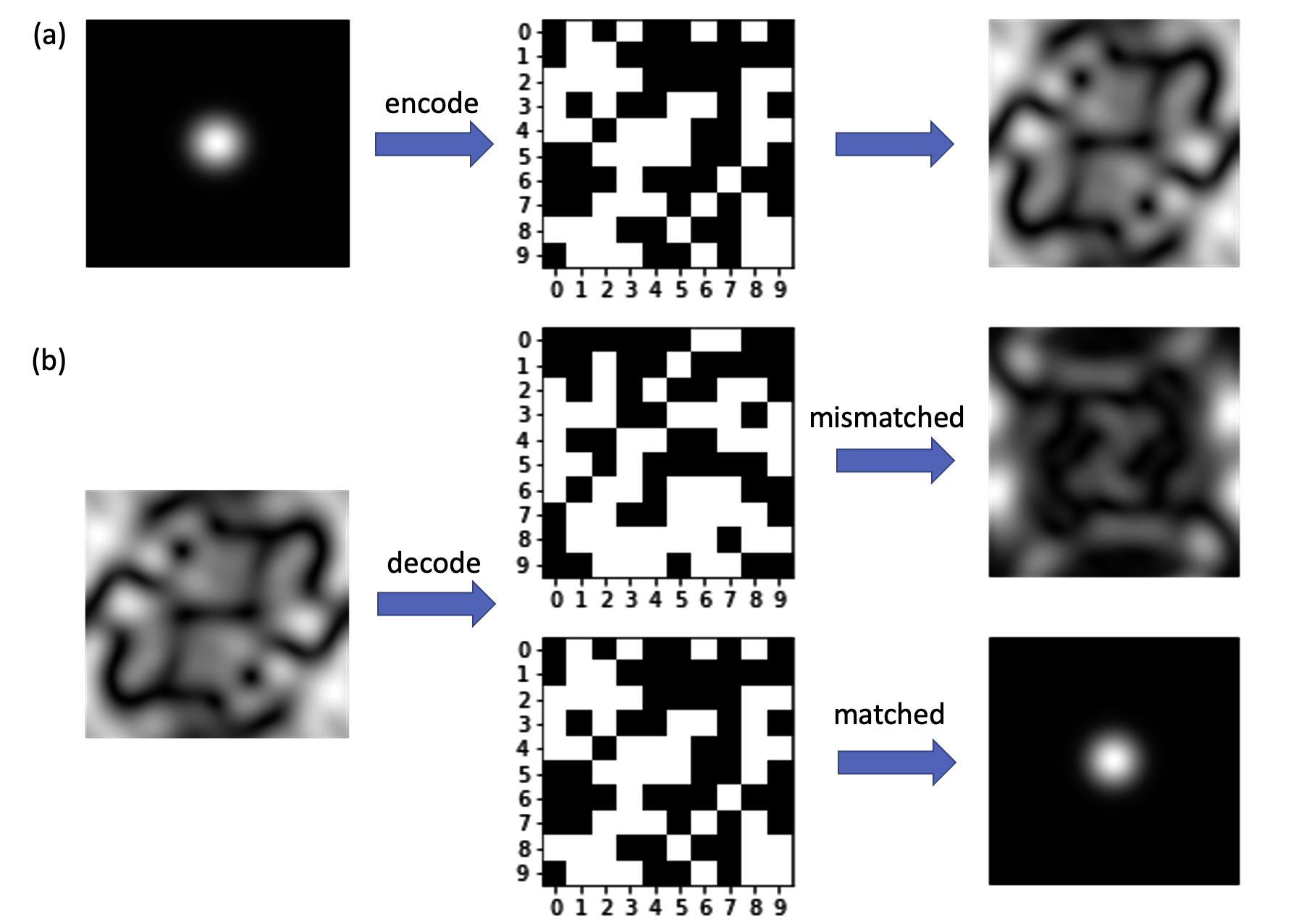}
\caption{Principle of orthogonal spatial coding. (a) Encoding an image. A random matrix of 0 and $\pi$ phase shifts are applied to the Fourier transform of the image sent to the receiver. The encoding phase mask introduces artificial aberrations to the image and produces a distorted and blurry image at the receiver's end. For clarity and simplicity, the image sent here is a simple Gaussian beam and the size of the encoding matrix is chosen to be $10\times\ 10$. (b) Decoding the transferred pattern. The Gaussian beam is recovered when the receiver applies a matching phase mask. Any other phase masks keep the image distorted and blurry. }
\label{fig:coding}
\end{figure}

In our experiment, the spatial code is generated by a random phase mask on the SLM placed at the Fourier plane of the seed beam.We first create a $N \times N$ matrix ($N < 1000$) where each entry $\phi^{rand}_{ij}$ takes either 0 or $\pi$. This gives the spatial code we use to encode the image transferred to the receiver (Figure \ref{fig:coding}). Then we expand the $N \times N$ matrix to the entire SLM window using a bicubic interpolation method \cite{bicubic} to ensure the smooth phase change across the phase mask printed on the SLM. This resulting phase profile $\phi^{rand}$ is used on both SLMs to introduce and cancel the effect of the spatial coding in the system. The decoding phase mask printed on the SLM in the idler beam should be the same as the phase mask on the SLM that blurs the image on the seed beam, except for the fact that it is rotated 180 degrees. This centro-symmetry is due to the momentum anticorrelation between the idler and the seed.

\begin{figure}[hbt!]
  \centering
  \includegraphics[width=0.5\textwidth]{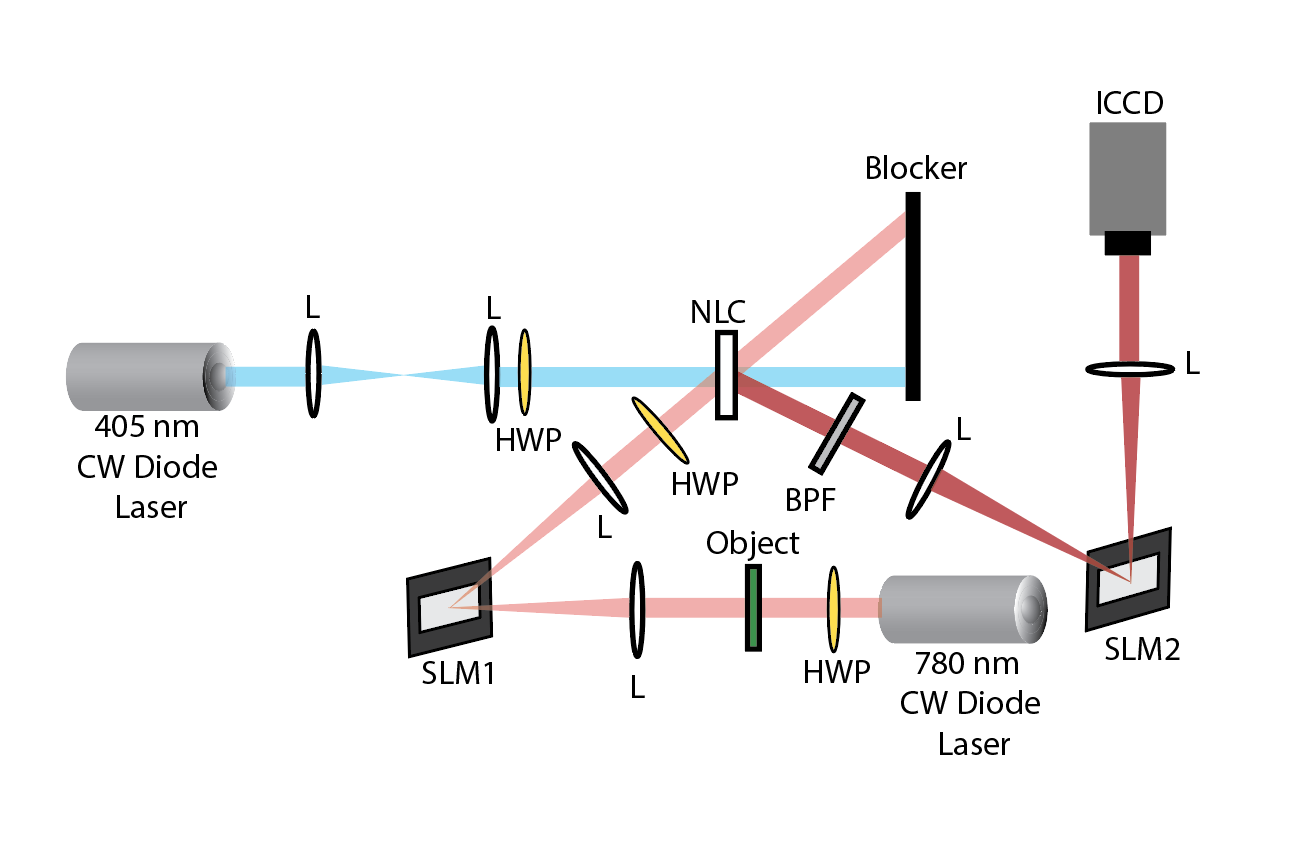}
\caption{Schematic of the experimental setup. L: lens; HWP: half-wave plate; BPF: band-pass filter; NLC: Type-II nonlinear crystal (BBO). A Type-II BBO crystal is pumped by a 405nm CW laser beam with a Gaussian spatial profile (drawn in blue). SLM1 generates a hologram that encodes the image in the Fourier plane of the object illuminated by the signal beam (drawn in pink). The encrypted image, which is imaged onto the BBO crystal, seeds the DFG process. The phase mask on SLM2 recovers the image transferred to the idler beam (drawn in red). }
\label{fig:setup}
\end{figure}

Figure \ref{fig:setup} shows the experimental setup used to implement the orthogonal spatial coding of images with DFG. A Type-II beta barium borate (BBO) nonlinear crystal was pumped by a horizontally polarized, 20 mW, c.w. 405 nm collimated Gaussian beam with a diameter of 1.7 mm produced by a laser diode. The seed beam with a wavelength of 780 nm was first collimated and then illuminated an object that features the USAF 1951 target. Both the pump and the seed beam were spectrally filtered with a narrowband filter (10 nm) centered at 405 nm and 780 nm respectively. A half-wave plate was placed after each narrowband spectral filter to control the polarization of the pump and the signal beam so that the phase-matching condition of the nonlinear frequency conversion is satisfied. The object, illuminated by the 780 nm seed beam, was imaged onto the Type-II BBO nonlinear crystal by a $4f$ ($f = 20$ cm) image-relay system. A spatial light modulator (Meadowlark E-series 1920 $\times$ 1200 SLM) was used to introduce the encryption key to the image in the seed beam. The image is encoded by the random phase mask printed on the SLM placed in the Fourier plane of the seed beam. The DFG process between the pump beam and seed beam is observed when the phase-matching condition is satisfied. The encoded conjugate image is transferred to the idler beam whose wavelength is 842nm. Another $4f$ system ($f = 20$cm) relays the conjugate image produced in the DFG onto an ICCD camera. A second spatial light modulator (Santec SLM-200) was placed at the Fourier plane of the idler beam to decode the transferred conjugate image. 

\begin{figure}[hbt!]
\centering
\includegraphics[width=0.5\textwidth]{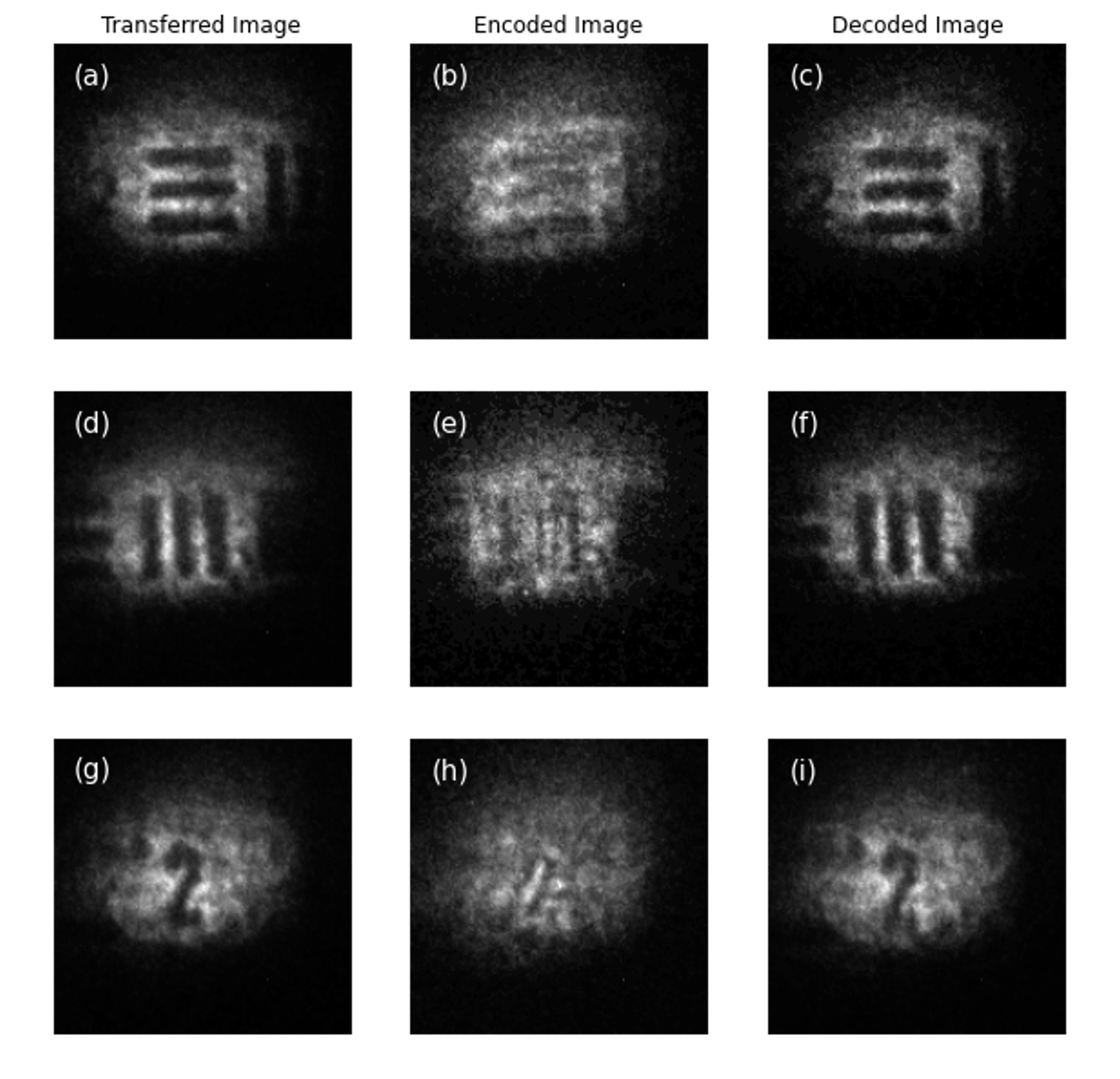}
\caption{Images captured by the ICCD camera. (a-c): original image, distorted image, and the decoded image of three horizontal bars transferred to the idler beam at 842nm; (d-f): original image, distorted image, and the decoded image of three vertical bars transferred to the idler beam at 842nm; (g-i): original image, distorted image, and the decoded image of "2" transferred to the idler beam at 842nm.}
\label{fig:result}
\end{figure}

We transferred three different objects, three horizontal bars, three vertical bars, and the number "2" on the USAF 1951 target with our image coding setup. The down-converted idler beam spans around 2.4mm in width and around 1.6mm in height. Figure \ref{fig:result}a, \ref{fig:result}d, and \ref{fig:result}g show the images of the objects transferred onto the 842nm idler beam when the encoding phase mask is turned off. Figure \ref{fig:result}b, \ref{fig:result}e, and \ref{fig:result}h show the distorted images collected on the ICCD camera when a random phase mask, $\phi_i(\mathbf{k}_i)$, is applied to the image in the seed beam. The phase value on the SLM in the idler arm is set to zero everywhere so that no decoding is made. Figure \ref{fig:result}c, \ref{fig:result}f and \ref{fig:result}i show the recovered images when the correct decoding phase mask, $\phi_i(\mathbf{k}_i) = \phi_s(-\mathbf{k}_s)$, is printed on the SLM in the idler arm. In all the images collected by ICCD camera, we can observe some minor discrepancies from the original USAF 1951 target images, even in the transferred images without encryption. This is because the image transfer in the nonlinear crystal is highly sensitive to its position. A perfect transfer of the spatial structure to the idler requires a spatial overlap of the seed and the pump structures exactly at the spot where down-conversion takes place in the nonlinear crystal. Overall, the results shown in Fig\ref{fig:result} indicate that the coded image can be effectively transferred and restored through a non-degenerate stimulated parametric down-conversion process as expected. 

In our orthogonal spatial coding scheme, the color of the received image is different from the image sent out in the seed beam. Therefore, we expect that, with proper modifications in our setup, our image coding setup may also find a suitable place in many frequency conversion imaging schemes\cite{Barh:19}. In these experiments, the photon interacting with the sample is up or down converted into a different wavelength before it is collected by the detector because direct observation of samples under UV or visible light often induces autofluorescence\cite{wang2010upconversion}, and a long exposure of the biological samples to high-frequency electromagnetic radiation can also cause sample photodamage and phototoxicity\cite{wu2015upconversion}. The two-color aberration cancellation technique used in this letter may suggest new possibilities to correct image distortions caused by anisotropy and optical defects in biological samples in many frequency conversion imaging applications. 

In conclusion, we have proposed and demonstrated an orthogonal spatial coding scheme based on the classical difference frequency generation. In the low-gain regime, the output field of the stimulated parametric down-conversion is proportional to the product of the pump field and the conjugate of the seed field. Making use of this property, our experiment has shown that the encoded image in the seed beam can be transferred to the idler beam at a different wavelength. The distorted image transferred to the idler beam can be recovered by canceling the effect of the spatial encryption with a proper corrective phase mask placed at the Fourier plane of the idler. We expect that the scheme proposed in this letter can be further developed into several practical frequency-conversion imaging schemes in aberrant media.

\bibliography{prl}

\end{document}